\documentclass[aps,prb,twocolumn,superscriptaddress,floatfix,showpacs]{revtex4}
\usepackage{textcomp}
\usepackage{amsmath}
\usepackage{amssymb}
\usepackage{graphicx,graphics}
\usepackage{latexsym,verbatim}
\usepackage{dcolumn} 
\usepackage{color}
\usepackage{cancel}
\usepackage{ulem}
\usepackage[latin1]{inputenc}

\begin{document}

\title{Chiral Superconductivity in Thin Films of doped Bi$_2$Se$_3$}

\author{Luca Chirolli}
\affiliation{IMDEA-Nanoscience, Calle de Faraday 9, E-28049 Madrid, Spain}
\email{luca.chirolli@imdea.org}

\begin{abstract}

Recent experimental evidences point to rotation symmetry breaking superconductivity in doped Bi$_2$Se$_3$, where the 
relevant order parameter belongs to a two-component odd-parity representation $E_u$ of the crystal point group. The $E_u$ 
channel admits two possible phases, the nematic phase, that well explains the reported rotation symmetry breaking, and the 
chiral phase, that breaks time-reversal symmetry. In weakly anisotropic three-dimensional systems the nematic phase is the stable one. 
We study the stability of the nematic phase versus the chiral phase as a function of the anisotropy of the Fermi surface and the thickness 
of the sample and show that by increasing the two-dimensional character of the Fermi surface or by reducing the number of layers in 
thin slabs the chiral phases is favoured. For the extreme 2D limit composed by a single layer of Bi$_2$Se$_3$ the chiral phase is 
always the stable one and the system hosts two chiral Majorana modes flowing at the boundary of the system. 

\end{abstract}

\pacs{74.20.Rp, 74.20.Mn, 74.90.+n}

\maketitle

\section{Introduction}
Chiral superconductivity is a topological quantum state of matter in which an unconventional superconductor spontaneously breaks 
time-reversal symmetry and develops an intrinsic angular momentum \cite{SigristRMP1991}. Its peculiar gap structure realizes a 
triplet $p_x+ip_y$ state that is topologically non-trivial. Key signatures in two-dimensional (2D) systems 
are chiral Majorana edge modes and Majorana zero energy states in vortex cores \cite{ZhangRMP2011,ReadGreen2000,IvanovPRL2001,
Alicea,Beenakker2013rev,Aguado}. In three dimensions, chiral superconductivity (SC) is also possible, allowing the realization of a Weyl 
superconductor with Majorana arcs on the surface \cite{MengPRB2012,SauPRB2012,YangPRL2014}. A possible candidate material for 
hosting this superconducting state is SrPtAs \cite{Biswas2013,Fisher2014}. Chiral superconductors have attracted great interest for their 
unconventional character and their potential use in the field of quantum computation \cite{NayakRMP2008,DasSarma2015}.

Recently, strong experimental evidences of unconventional superconductivity have been reported for a well known material, Bi$_2$Se$_3$, 
that in its pristine form is a Topological Insulator (TI) \cite{ZhangNP2009,HasanKaneRMP2010}. 
Possibly odd-parity superconductivity was first reported in Bi$_2$Se$_3$ intercalated with Cu\cite{HorPRL2010,Wray2010,KrienerPRL2011}, 
although clear evidence for the characteristic surface Andreev states has remained controversial \cite{SasakiPRL2011,LevyPRL2013,PengPRB2013}. 
The first studies motivated the theoretical characterization of three dimensional, time-reversal invariant (TRI) topological superconductivity in centrosymmetric 
systems\cite{FuBerg}.  A much richer phenomenology has recently emerged, showing a broken $C_3$ symmetry in the superconducting state in samples 
intercalated with Cu, Nb, and Sr\cite{shrutiPRB2015,liuJACS2015,wangCM2016,asabaPRX2017}. Several experiments reported uniaxial anisotropy 
response to an in-plane magnetic field in the Knight shift\cite{Matano2016}, the upper critical field\cite{YonezawaNP2016,
PanSR2016}, the magnetic torque\cite{asabaPRX2017}, and the specific heat\cite{YonezawaNP2016}. Specific heat\cite{KrienerPRL2011} and penetration 
depth\cite{SmyliePRB2016,Shen2017} have excluded the presence of line nodes on the Fermi surface. All these observations support a pairing 
state of nematic type belonging to the two-component representation $E_u$ of the crystal point group\cite{FuPRBR2014,Venderbos2015}.

Theoretical modelling have also discussed different aspects of the $E_u$ states, covering from bulk properties\cite{HashimotoJPSJ2013,
NagaiPRB2016,VenderbosPRB2016-2}, to surface states\cite{YangPRL2014}, vortex states\cite{FengchengPRB2017-1,ZyuzinPRL2017}, 
the interplay between $E_u$ superconductivity and magnetism in promoting time-reversal symmetry breaking 
states\cite{ChirolliPRB2017,NoahPRB2017}, and the role of odd-parity fluctuations as the mechanism at the basis of $E_u$ superconductivity
\cite{FengchengPRB2017} and preemptive nematicity abouve $T_c$ \cite{Fernandes2012,Schmalian2017}.

In this work we study superconductivity in Bi$_2$Se$_3$ in the $E_u$ odd-parity channel, focusing on the stability of the nematic phase versus 
the chiral phase as a function of the anisotropy of the system and the thickness of the sample. Bi$_2$Se$_3$ is a layered material in which the 
unit cell is constituted by a Quintuple Layer (QL) structure. It is therefore reasonable to study the behaviour of the system by varying the interlayer coupling 
and the chemical potential. We show that an increase of the two-dimensional character of the Fermi surface favours the chiral phase. Chemical dopants intercalate 
between the unit cells and modify their distance and relative coupling, together with the charge density. Strong anisotropy can be achieved by increasing 
the doping or by properly choosing the dopants so to increase the interlayer spacing of the materials. 

Interestingly, a second root towards chiral 
superconductivity is provided by exfoliation. In particular, the chiral phase is the natural phase of the $E_u$ channel in the extreme 2D limit of a 
single layer \cite{FuPRBR2014,Venderbos2015}. We show that by reducing the thickness of the sample without increasing the anisotropy of the 
system naturally drives the system towards the chiral phase. We find as a rough estimate that a thin slabs with approximately ten layers marks 
the stability threshold between the nematic and the chiral phase. Experimentally, exfoliation down to the single QL case has been achieved
\cite{ZhangNP2010crossover,ZhangAFM2011}, making this root a promising way toward chiral quasi-2D superconductors.

The single layer case acquires high relevance in the context of 2D materials engineering, 
whereby properties of a material can be fine tuned by coupling with a proper substrate. By placing a single layers of Bi$_2$Se$_3$ on top of a 
suitable substrate, planar mirror inversion symmetry breaks explicitly, the point group is reduced to $C_{3v}$, and the system is expected to show 
Rashba spin-orbit interaction. This possibility becomes highly relevant in the light of recent theoretical developments concerning time-reversal 
symmetry breaking in 2D non-centrosymmetric systems \cite{Scheurer2017}, according to which the superconducting state can break time-reversal 
symmetry only in presence of a threefold rotation symmetry.  As shown in Ref.~[\onlinecite{Scheurer2017}], if the superconducting order parameter belongs to the $E_u$ representation the chiral state must appear for sufficiently 
large Rashba coupling. This implies that the SC order parameter only gradually changes as the surface Rashba coupling is included, even if the 
Kramers degeneracy is lifted. These considerations boost single layers of Bi$_2$Se$_3$ as an optimal candidate for the observation of chiral 
superconductivity in 2D systems. 

The realization of the chiral state promotes the system to class D topological superconductors that in 2D are characterized by a ${\mathbb Z}$ 
topological invariant and are expected to show chiral Majorana modes flowing at the boundary \cite{Schnyder}. The number of chiral Majorana modes is 
dictated by the Chern number, that in the present case takes the value $C_{\rm ch}=\pm 2$ for the $p_x\mp ip_y$ solution. 
Starting from a tight-binding model that well approximates the complicated band structure 
of Bi$_2$Se$_3$, we show that the chiral phase in this material supports two chiral Majorana modes that copropagate at the boundary of the 
system and can find useful applications in interferometric schemes\cite{Chirolli2018}. The low energy Hamiltonian of the system is a massive 
Dirac Hamiltonian, so that our results apply to generic systems that share the same low energy description, such as TI thin films \cite{Parhizgar2017} 
and Rashba bilayer system \cite{Nakosai-PRL2012}.

The work is structured as follows: in Sec.~\ref{Sec:EuSC} we review the known analysis of the two-component superconducting channel of 
the $D_{3d}$ crystal point group. In Sec.~\ref{Sec:GL} we derive the Ginzburg Landau function that describes the condensation of the 
two-component channel. In Sec.~\ref{Sec:Chiral} we study the stability of the chiral phase and show that in the strong anisotropic case it is the 
favoured phase. In Sec.~\ref{Sec:Chiral-vs-N} we show that by reducing the thickness of the sample a chiral phase is obtained for thin slabs. 
In Sec.~\ref{Sec:Topo} we study the surface states through a tight-binding numerical simulation. Finally, in Sec.~\ref{Sec:conclusions} we 
conclude with a summary of the results.

\section{$E_u$ Superconductivity}
\label{Sec:EuSC}

We consider doped  Bi$_2$Se$_3$ in the ${\it k}\cdot{\it p}$ low energy approximation introduced in Ref.~[\onlinecite{FuBerg}]. The point group of the 
crystal is $D_{3d}$ and the system can be described by a simplified model in which the unit cell is constituted by a bilayer 
structure where spin $s$ electrons occupy $p_z$-like orbitals on the top (T) and bottom (B) layers of the microscopic QL unit cell. 
The low energy Hamiltonian is described by a massive anisotropic ($v_z\neq v$) Dirac model that reads
\begin{equation}\label{Eq:Hdirac}
H^0_{\bf k} = m\sigma_x + v(k_xs_y-k_ys_x)\sigma_z+v_zk_z\sigma_y,
\end{equation}
where  Pauli matrices $\sigma_i$ and $s_i$ describe the orbital and spin degrees of freedom, respectively. The Hamiltonian is TRI, where the 
time reversal operator is ${\cal T} = is_y K$ with $K$ complex conjugation.

Superconductivity is described within the Bogoliubov deGennes (BdG) formalism by introducing the Nambu spinor 
$\Psi_{\bf k} = ({\bf c}_{\bf k}, i s_y {\bf c}_{-{\bf k}}^\dag)^T$, with ${\bf c}_{\bf k}$ fermionic annihilation operators 
of the Hamiltonian $H_{\bf k}^0$. The Hamiltonian reads 
$\hat{\cal H}=\frac{1}{2}\int d{\bf k}\Psi^\dag_{\bf k} H({\bf k}) \Psi_{\bf k}$, with
\begin{equation}\label{Eq:BdG}
H_{\bf k} = (H^0_{\bf k}-\mu)\tau_z + \Delta_{\bf k} \tau_+ + \Delta^\dag_{\bf k} \tau_-,
\end{equation}
and with $\Delta_{\bf k}$ generic momentum-dependent $4\times 4$ gap matrices. The Nambu construction imposes that $H_{\bf k}$ has a 
charge conjugation symmetry ${\cal C}$ implemented as $U_{\cal C} H(-{\bf k})^* U_{\cal C}^\dagger = -H({\bf k})$, with $U_{\cal C} = s_y \tau_y$. 
${\cal C}$ imposes a restriction on the pairing matrix, $s_y\Delta^*(-{\bf k})s_y = \Delta({\bf k})$. If pairing is momentum independent, there are only 6 
possible matrices in the irreducible representations of the $D_{3d}$ point group that satisfy this constraint and they have been classified in 
Ref.~[\onlinecite{FuBerg}]. Accordingly, they are given by the even parity channel $I$ and $\sigma_x$ belonging to the $A_{1g}$ representation, 
the odd-parity channel $\sigma_ys_z$ belonging to $A_{1u}$ representation,  $\sigma_z$ belonging to $A_{2u}$ representation, and 
$(-\sigma_y s_y,\sigma_y s_x)$ belonging to $E_u$ representation. In particular, the latter forms a two-component representation that can 
describe nematic or chiral SC \cite{FuPRBR2014,Venderbos2015}.

Focusing on the $E_u$ odd-parity channel we associate to the matrix operators the following order parameters
\begin{equation}
\boldsymbol{\psi}=(\psi_x,\psi_y) \sim (-\sigma_y s_y,\sigma_y s_x) \sim E_u\nonumber.
\end{equation}
In Ref.~[\onlinecite{FuBerg}] it was shown that when only local pairing is considered, the $A_{1u}$ is the leading instability in a wide 
range of parameters in the phase diagram. On the other hand, the author has shown that inclusion of momentum-dependent pairing  
terms only affects the critical temperature of the nematic channel $E_u$ \cite{ChirolliPRB2017}, rising it with respect to the critical 
temperature of the $A_{1u}$ channel. Recently, odd-parity fluctuations together with repulsive Coulomb interactions have also emerged 
as a possible mechanism that selects the $E_u$ odd-parity two-component channel as the leading SC channel \cite{FengchengPRB2017}. 
We then assume that the nematic channel condenses and focus on the competition between the nematic and chiral phases.

\section{Ginzburg Landau Theory}
\label{Sec:GL}

We start considering the $E_u$ phase $\boldsymbol{\psi}$ and study the conditions under which a chiral phase occurs. Symmetry dictates 
the form of its free energy that reads
\begin{eqnarray}\label{Eq:GLpsi}
F_\psi &=& a |\boldsymbol{\psi}|^2 + b_1 |\boldsymbol{\psi} |^4  + b_2 |\psi_x\psi_y^*-\psi_y\psi^*_x|^2.
\end{eqnarray}
The $E_u$ representation admits two possible superconducting states: a nematic state $\boldsymbol{\psi} \propto (1,0)$ which is time-reversal 
invariant and has point nodes on the equator of the Fermi surface, and a chiral state $\boldsymbol{\psi} \propto (1,i)$ which breaks TR symmetry 
and has $C=\pm 2$ Weyl nodes at the north and south pole of the Fermi surface\cite{Venderbos2015}. The sign of the coupling $b_2$ determines 
whether the $E_u$ representation chooses the nematic (for $b_2>0$) or the chiral state (for $b_2<0$). Microscopic calculations show that for a 3D 
isotropic model $b_2>0$, so that no TRB phase may arise in the system \cite{FuPRBR2014,Venderbos2015}. We now specifically study the sign of 
the coupling $b_2$ versus Fermi surface anisotropy and sample thickness. 

Setting the chemical potential in the conduction band, $\mu>m$, we can reduce the dimensionality of the problem by projecting the 
Hamiltonian and the gap matrix down to the conduction band, so that the gap matrix reads
\begin{equation}
\Delta_{\bf k}=\psi_x{\bf d}_x\cdot\tilde{\bf s}+\psi_y{\bf d}_y\cdot\tilde{\bf s},
\end{equation}
where ${\bf d}_x=(0,-\tilde{k}_z,\tilde{k}_y)$ and ${\bf d}_y=(\tilde{k}_z,0,-\tilde{k}_x)$, 
the momentum has been rescaled as $\tilde{\bf k}=(vk_x,vk_y,v_zk_z)/\mu$,  and  $\tilde{\bf s}$ is a momentum-dependent spin-1/2 like 
vector operator parametrizing the twofold degenerate subspace at every ${\bf k}$ point associated to Kramers degeneracy \cite{FuPRL2015}. 
Explicitly, defining $|\psi_{{\bf k},1}\rangle$ and $|\psi_{{\bf k},2}\rangle$ the two degenerate eigenstates in the conduction band at momentum ${\bf k}$, 
the vector $\tilde{\bf s}$ is obtained as $\tilde{s}_x=|\psi_{1,{\bf k}}\rangle\langle\psi_{2,{\bf k}}|+{\rm H.c.}$, 
$\tilde{s}_y=-i|\psi_{1,{\bf k}}\rangle\langle\psi_{2,{\bf k}}|+{\rm H.c.}$, and  
$\tilde{s}_z=|\psi_{1,{\bf k}}\rangle\langle\psi_{1,{\bf k}}|-|\psi_{2,{\bf k}}\rangle\langle\psi_{2,{\bf k}}|$.

We can now integrate away the fermionic degrees of freedom and obtain a non-linear functional for the order parameters
\begin{equation}
{\cal S}=\int_0^\beta d\tau\frac{1}{V}{\rm Tr}\left[\hat\Delta^\dag\hat\Delta\right]-\frac{1}{\beta}{\rm Tr}\ln(-{\cal G}_0^{-1}+\Sigma),
\end{equation}
with $-{\cal G}_0^{-1}=\partial_\tau+(H_0-\mu)\tau_z$, $\Sigma=\tau_+\hat{\Delta}+\tau_-\hat{\Delta}^\dag$, $\beta=1/T$ the inverse temperature, 
and the trace is over all the degrees of freedom, ${\rm Tr}\equiv T\sum_\omega\int d{\bf k}$. As usual, the microscopic GL theory is obtained by 
expanding the non-linear action in powers of the fields,
\begin{equation}
{\rm Tr}\ln(-{\cal G}_0^{-1}+\Sigma)={\rm Tr}\ln(-{\cal G}_0^{-1})-\sum_{n=1}^\infty\frac{1}{n}{\rm Tr}({\cal G}_0\Sigma)^n.
\end{equation}
The forth order coefficient are determined by the forth order averages $\langle\Delta_{\bf k}\Delta_{\bf k}^*\Delta_{\bf k}\Delta_{\bf k}^*\rangle$, 
where $\langle\ldots \rangle=T\sum_{\omega_n}\int \frac{d{\bf k}}{(2\pi)3}G_+^2G_-^2{\rm Tr}[\ldots]$, $G_\pm=(i\omega_n\mp\xi_{\bf k})^{-1}$, 
$\xi_{\bf k}=\epsilon_{\bf k}-\mu$ and $\epsilon_{\bf k}=\sqrt{\mu^2\tilde{\bf k}^2+m^2}$ is the dispersion of the conduction band. 
Explicitly, the fourth order terms are given by
\begin{eqnarray}
b_1&=&3\langle ({\bf d}_x\cdot{\bf d}_y)^2\rangle+\langle ({\bf d}_x\times{\bf d}_y)^2\rangle,\\
b_2&=&-\langle ({\bf d}_x\cdot{\bf d}_y)^2\rangle+\langle ({\bf d}_x\times{\bf d}_y)^2\rangle.
\end{eqnarray}
Clearly, parallel vectors ${\bf d}_i$ favor a chiral phase and orthogonal vectors favor a nematic 
phase.

\section{Chiral phase for strong anisotropy}
\label{Sec:Chiral}

We now study the parameter $b_2$ as a function of the anisotropy of the Fermi surface.    
By performing the averages one can approximate
\begin{equation}\label{Eq:b2}
b_2=\kappa \int d^3\tilde{k}\left[\tilde{k}_{z}^2 (\tilde{k}^2+\tilde{k}_{z}^2)-\tilde{k}^4/8\right]\delta(\xi_{\tilde{\bf k}}/\mu),
\end{equation}
where $\kappa=7\zeta(3)N_F/(8(\pi T_c)^2)$, $N_F=\int d^3{\bf k}\delta(\xi_{\bf k})/(2\pi)^3$ the density of states at the Fermi level, and $\tilde{k}^2=\tilde{k}^2_x+\tilde{k}^2_y$. 
For an isotropic Fermi surface the coefficient $b_2$ is positive and the nematic phase is favoured. By inspection of Eq.~(\ref{Eq:b2}) 
it becomes clear how a strong anisotropy of the Fermi surface can drive the system into the $b_2<0$ regime. 

\begin{figure}[t]
\begin{center}
\includegraphics[width=0.45\textwidth]{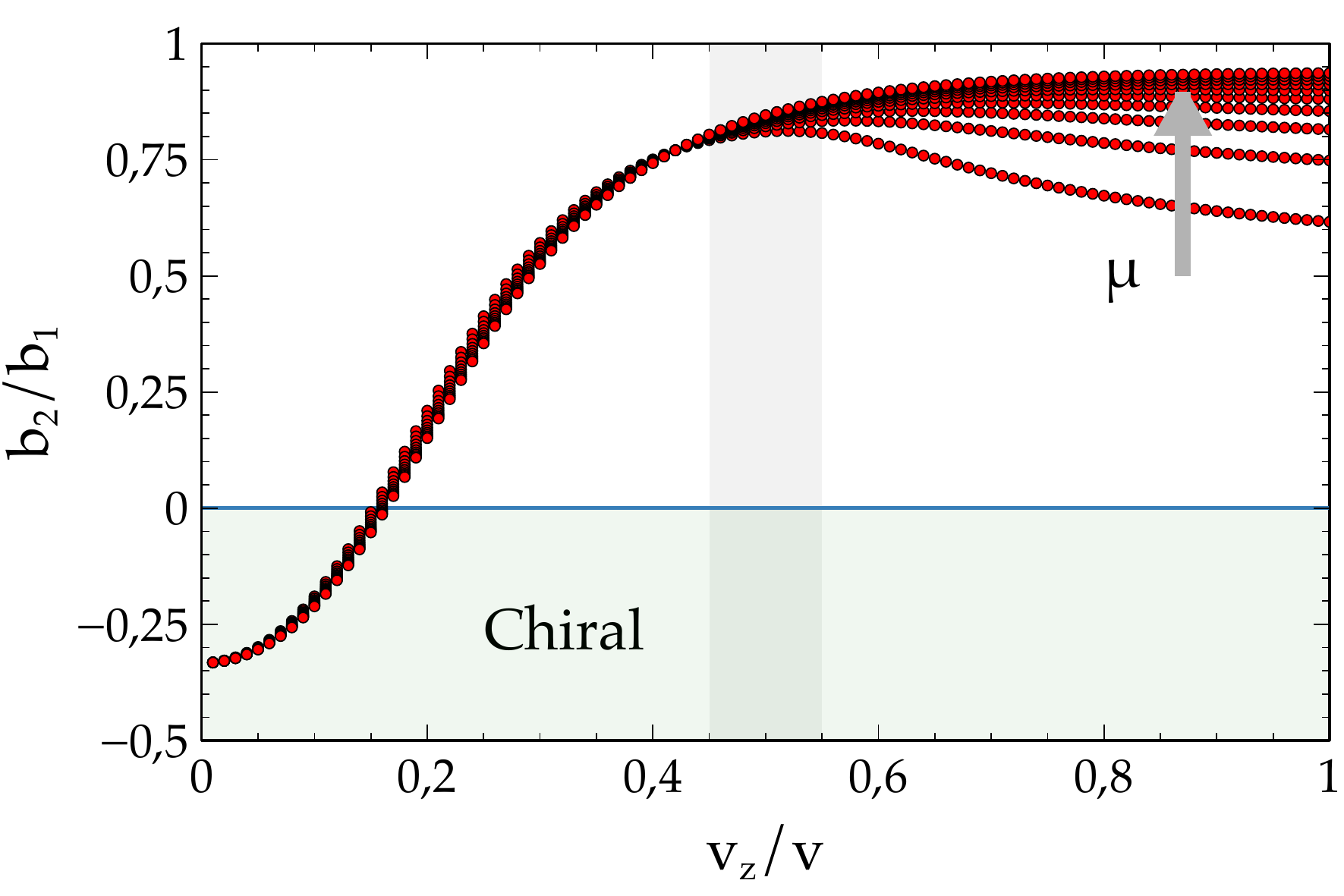}
\caption{(color online) Variation of the coefficient ratio $b_2/b_1$ as a function of the anisotropy of the system parametrized 
by $v_z/v$ and the chemical potential $\mu$, keeping the parameter $\alpha_z$ fixed. For sufficiently small $v_z/v$ the 
coefficient $b_2$ becomes negative ($b_1>0$) and the chiral phase becomes possible. \label{Fig:b2}}
\end{center}
\end{figure}

The Hamiltonian Eq.~(\ref{Eq:Hdirac}) is linear in momentum and characteristic surface states of the TI arise when ${\rm sign}(m v_z)<0$ for 
states confined in $z>0$ \cite{HsiehFu-PRL2012}. Nevertheless, quadratic corrections in the mass term can be also considered and appear in 
more refined band structure calculations\cite{ZhangNP2009},
\begin{equation}\label{Eq:mass-k}
m({\bf k})=m-\alpha k^2-\alpha_z k_z^2,
\end{equation}
where $k^2=k_x^2+k_y^2$. For $m,\alpha,\alpha_z>0$ the mass term changes sign on a particular surface in momentum space. 
This property yields a non-trivial topology of the insulator. For simplicity we neglect a spin- and orbital-independent term that 
adds to the Hamiltonian as a diagonal ${\bf k}$-dependent contribution and does not change the topological properties of the system, 
a part from breaking the particle-hole symmetry of the Dirac Hamiltonian describing the topological insulator. 

The momentum dependence of the mass term introduces a second scales $\alpha_z$ along the $k_z$ direction that, together with 
the velocity $v_z$, makes the Fermi surface intrinsically anisotropic. If $\alpha_z$ is neglected, the unique scale $v_z$ can be 
reabsorbed in a redefinition of the momentum and it eventually factorizes in the expression of $b_1$ and $b_2$, in a way that 
their value become fixed and positive. It is then reasonable to study the parameter $b_2$ as we increase the anisotropy of the 
Fermi surface by considering finite $\alpha_z$.

The values of $v_z$ and $\alpha_z$ can be controlled by chemical doping, in that dopants intercalate between the QLs and modify 
the interlayer distance $a_z$ and hopping $t_{z}$. The latter can be assumed to be exponentially dependent on $a_z$ itself, 
$t_{z}=t_{z}^0\exp(-a_z/R)$, with $R$ a microscopic length scale characteristic of the $p_z$ orbital of Se, and $t_{z}^0$ the 
amplitude of the hopping integral. It then follows that an increase in the doping is expected to lower both $v_z$ and $\alpha_z$.  

In Fig.~\ref{Fig:b2} we plot the dependence of $b_2/b_1$ as a function of $v_z/v$, keeping $\alpha_z$ constant and taking for 
reference the parameters of the well known model of Ref.~[\onlinecite{ZhangNP2009}], $m=0.28~$eV, $\alpha=56.6~$eV~\AA$^2$, 
$\alpha_z=10.0~$eV~\AA$^2$, and $v=4.1~$eV~\AA. The coefficient $\alpha_z$ drops from the ratio. The shadowed regions indicate 
the region around $v_z=2.2~$eV~\AA, that is realized in the undoped material. We clearly see that by decreasing $v_z/v$ we can obtain 
negative $b_2$ values. The strong topological character of the material allows a wide range of variation of $v_z$ through doping, without 
changing the topological nature of the system, so that a chiral phase can be obtained by properly choosing the dopants and their amount.

\section{Chiral phase in thin slabs}
\label{Sec:Chiral-vs-N}

In the 2D limit the Fermi surface is a line at $k_z=0$, the coefficient $b_2$ Eq.~(\ref{Eq:b2}) is explicitly negative, and the chiral phase is stable. 
The case $v_z\to 0$ and $\alpha_z\to 0$ is clearly realized when the system approaches the limit of quasi decoupled layers or the quasi 
2D limit, characterized by an open Fermi surface\cite{LahoudPRB2013}. We now study the stability of the nematic versus the chiral 
phase for reduced thickness of the sample by varying the number of unit cell layers. 

We consider a simplified tight-binding model along the lines of Ref.~[\onlinecite{HsiehFu-PRL2012}]. 
We approximate the QL structure as a bilayer system (BL) composed by its top most (T) and bottom 
most (B) Se layers, described by triangular lattices on top of each other. The entire structure is described 
by the tight-binding Hamiltonian
\begin{equation}\label{Htb-Dirac}
H_{\rm TI}=H_0+H_{\rm R}+H_z.
\end{equation}
The first term $H_0$ describes spin-independent hopping within the same 
layer and nearest neighbour tunnelling between the two layers 
\begin{eqnarray}
H_0&=&t_0\sum_{<i,j>,\sigma s}c^\dag_{i,\sigma,s}c_{j,\sigma,s}+t_1\sum_ic^\dag_{i,T,s}c_{i,B,s}+{\rm H.c.}\nonumber\\
&+&t_2\sum_{<i,j>,s}(c^\dag_{i,T,s}c_{j,B,s}+c^\dag_{i,B,s}c_{j,T,s}),
\end{eqnarray}
with $\sigma=T,B$ labeling the two layers. Atoms in the two layers experience local opposite electric fields along the $\pm \hat{z}$ 
direction, that give rise to Rashba SOI of opposite sign on the two layer in the form 
\begin{equation}
H_R=i\lambda\sum_{<ij>\sigma,ss'}p_\sigma c^\dag_{i,\sigma,s}c_{j,\sigma,s'}{\bf s}_{ss'}\cdot\hat{z}\times{\bf a}_{ij},
\end{equation}
with $p_{T,B}=\pm 1$ and ${\bf a}_{ij}$ a unit vector connecting site $i$ and site $j$. Finally, along the $z$-direction the  
structure is repeated as a series of tightly bound BL planes weakly coupled by van der Waals forces. Within the effective bilayer 
model the dynamics along the vertical direction can be described by an interlayer hopping term
\begin{equation}\label{Hz}
H_z=t_z\sum_jc^\dag_{j,T,s}c_{j-1,B,s}+{\rm H.c.},
\end{equation}
with $t_z$ intercell hopping amplitude.

\begin{figure}[t]
\begin{center}
\includegraphics[width=0.45\textwidth]{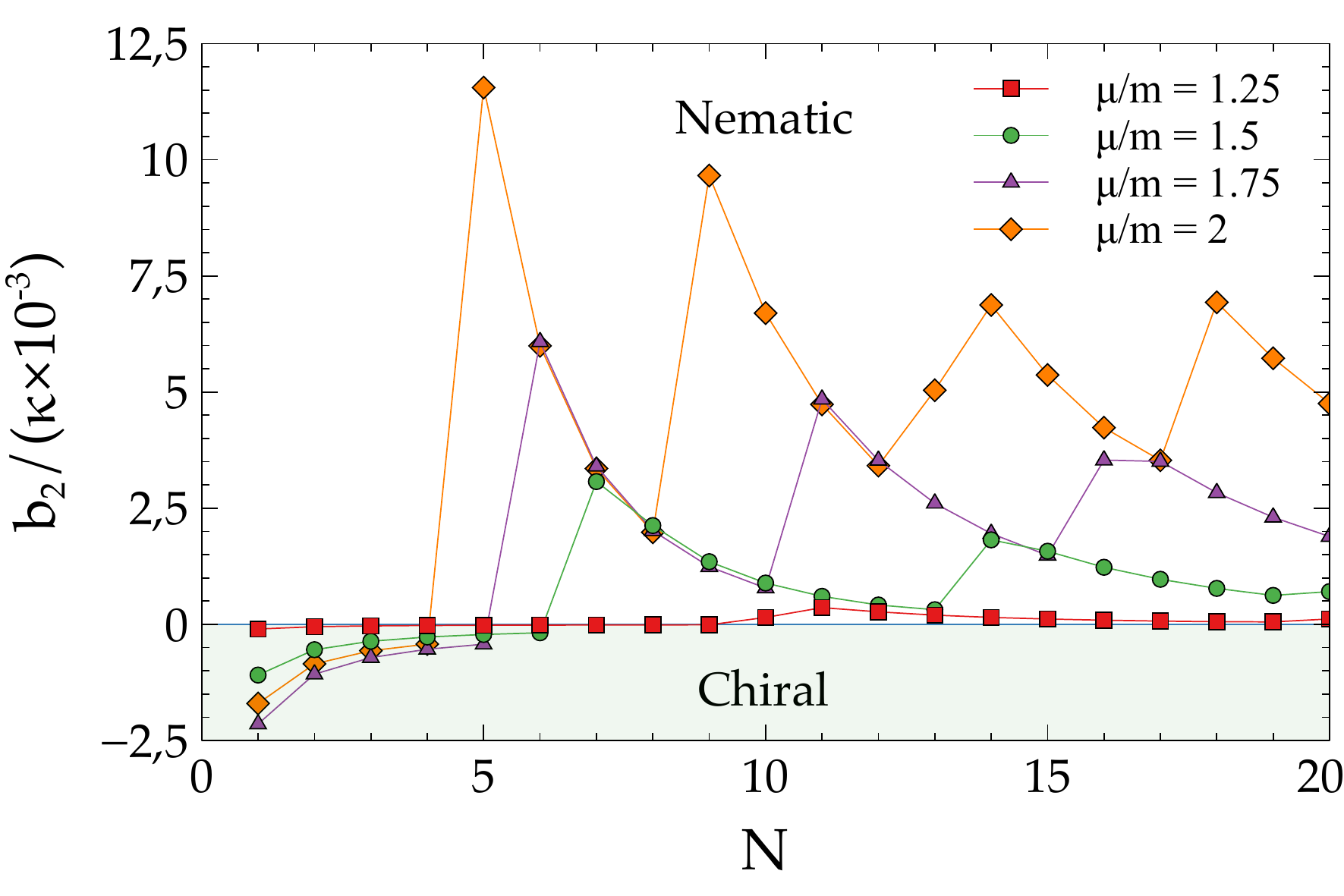}
\caption{(color online) Variation of the coefficient $b_2$ as a function of the number of layers $N$ for different chemical potential. 
The parameter used in the tight-binding model are $t_0=0$, $t_1=-4.1~{\rm eV}$, $t_2=0.75~{\rm eV}$, $t_z=-0.6~{\rm eV}$, 
$\lambda=0.5~{\rm eV}$, $a=5.0~{\rm \AA}$, and $a_z=9.1~{\rm \AA}$. \label{Fig:b2-vs-N}}
\end{center}
\end{figure}

The Hamiltonian $H_{\rm TI}$ describes a 3D TI, whose small momentum expansion well approximates the Hamiltonian 
Eq.~(\ref{Eq:Hdirac}) with in-plane velocity $v=3\lambda a$, out-of-plane velocity $v_z=t_{z}a_z$, and a mass term that 
depends on the momentum, $m({\bf k}) =6t_2+t_1+t_z-3t_2a^2k^2/2-t_{z}a_z^2/2$, from which we extract 
$m=6t_2+t_1+t_z$,  $\alpha=3t_2a^2/2$,  and $\alpha_z=t_{z}a_z^2/2$ appearing in Eq.~(\ref{Eq:mass-k}). 
In order to study the stability of the chiral phase for the massive Dirac model Eq.~(\ref{Eq:Hdirac}) we choose values of the 
tight-binding parameters $t_0$, $t_1$, $t_2$, $\lambda$, and $t_z$, in a way that the band structure is well described by a Dirac 
equation at small momentum. With the choice specified in Fig.~\ref{Fig:b2-vs-N} we obtain a mass $m=-0.2~{\rm eV}$ and 
velocities $v=7.5~{\rm eV}$\AA, $v_z=-5.46~{\rm eV}$\AA. These values are on order of those provided in Ref.~[\onlinecite{ZhangNP2009}] 
and the resulting model well describes the complicate band structure of Bi$_2$Se$_3$ at low energy.

The coefficient $b_2$ is calculated with the Fourier transformed Hamiltonian $H_{\rm TI}({\bf k}_\parallel,n_z)=
H_{\rm TI}({\bf k}_\parallel,\pi n_z/Na_z)$, where $N$ is the number of layers, and by collecting the relevant 
terms in the fourth order expansion projected onto the conduction band
\begin{equation}
F_4=\frac{\kappa}{N_F}\sum_{n_z,{\bf k}}\delta(\xi_{{\bf k},n_z}){\rm Tr}[\left({\cal P}_{{\bf k},n_z}\Delta^\dag {\cal P}_{{\bf k},n_z}\Delta\right)^2],
\end{equation}
where ${\cal P}_{{\bf k},n_z}=\sum_{i=1,2}|\psi_{{\bf k},i}\rangle\langle\psi_{{\bf k},i}|$ is the projection operator of the conduction 
band subspace and $\Delta=-\psi_x\sigma_y s_y+\psi_y\sigma_y s_x$. We keep the tight-binding parameters fixed and only vary 
the number of layers $N$. In Fig.~\ref{Fig:b2-vs-N} we clearly see that the chiral phase is stable for sufficiently thin slabs of material. 
In particular, we see that the coefficient $b_2$ experiences quantum oscillations due to the coupling between the 2D layers. As a 
function of the chemical potential, small negative values of $b_2$ are obtained already for thick slabs and low doping, whereas larger 
negative $b_2$ values require higher doping and  thinner slabs. 

The coefficient $b_2$ follows the dispersion along the $z$ direction. By reducing the number of layers the energy of the subbands 
grows and $b_2$ grows accordingly. As the energy of a given subband grows above the Fermi level the coefficient $b_2$ suddenly 
drops to the successive subband until a single band remains populated below the Fermi level. At that point $b_2$ becomes negative, 
as it cannot grow any longer. This explains why for low doping a structure composed by several unit cell develops the chiral phase, 
whereas for high doping one has to go down to the single layer case to encounter only one band below the Fermi level. 
The results of Fig.~\ref{Fig:b2-vs-N} are quite robust to variations of the tight-binding parameters characterizing the Hamiltonian, 
as long as the low energy model is well captured by a massive Dirac Hamiltonian. The threshold $N$ at which the 
transition to the chiral state takes place depends on the actual values of the tight-binding parameters. The quantum 
oscillations follow the band structure profile and they appear as long as the systems displays quantization of the subbands.

\section{Chiral Majorana Modes}
\label{Sec:Topo}

The results presented in the previous sections show that a feasible way of obtaining a chiral phase is through exfoliation and that the extreme 
case of a single layer is the best candidate for chiral superconductivity. The chiral solution is given by $\boldsymbol{\psi}=\psi_0(1,i)$ so that 
the resulting gap on the Fermi surface reads 
\begin{equation}\label{Eq:ChiralGap}
\Delta\propto \psi_0(k_y-ik_x)\tilde{s}_z, 
\end{equation}
where $\tilde{s}_z$ is the third Pauli matrix in the Kramers basis at momentum ${\bf k}$. Starting from the $k\cdot p$ Hamiltonian 
Eq.~(\ref{Eq:Hdirac}) at $k_z=0$ we can write the Kramer basis by employing the Manifestly Covariant Bloch Basis (MCBB) introduced 
in Refs.~[\onlinecite{FuPRL2015,Venderbos2015}], where the band eigenstates are chosen to be fully spin polarized along the $z$ direction 
at the origin of point group symmetry operations. 
\begin{equation}
|\psi_{{\bf k},1}\rangle=\left(\begin{array}{c}
\sqrt{\epsilon_k+m}\\
\sqrt{\epsilon_k+m}\\
ik_x-k_y\\
-ik_x+k_y
\end{array}\right),
|\psi_{{\bf k},2}\rangle=\left(\begin{array}{c}
-ik_x-k_y\\
ik_x+k_y\\
\sqrt{\epsilon_k+m}\\
\sqrt{\epsilon_k+m}\\
\end{array}\right),
\end{equation}
in the basis $(c_{{\bf k},T,\uparrow},c_{{\bf k},B,\uparrow},c_{{\bf k},T,\downarrow},c_{{\bf k},B,\downarrow})^T$, up to a normalization 
factor $1/(2\sqrt{\epsilon_k^2+m\epsilon_k})$. The action of time-reversal ${\cal T}=is_yK$, parity ${\cal P}=\sigma_x$, and mirror about 
$x$, $M_x=is_x$, are easily computed, resulting in the following transformation properties, ${\cal T}|\psi_{{\bf k},1}\rangle=-|\psi_{-{\bf k},
2}\rangle$, ${\cal T}|\psi_{{\bf k},2}\rangle=|\psi_{-{\bf k},1}\rangle$, ${\cal P}|\psi_{{\bf k},i}\rangle=|\psi_{-{\bf k},i}\rangle$, and 
$M_x|\psi_{{\bf k},1}\rangle=i|\psi_{M_x{\bf k},2}\rangle$, $M_x|\psi_{{\bf k},2}\rangle=i|\psi_{M_x{\bf k},1}\rangle$. It is then clear that 
$|\psi_{{\bf k},1}\rangle$ and $|\psi_{{\bf k},2}\rangle$ form a Kramers doublet and represent a preferential basis that transform like a 
spin-1/2 object \cite{FuPRL2015}. It is easily seen that the gap matrix $\Delta=\Delta_0(-\sigma_y s_y+i\sigma_y s_x)$ results in 
Eq.~(\ref{Eq:ChiralGap}) when projected on the basis spanned by $|\psi_{{\bf k},1}\rangle$ and $|\psi_{{\bf k},2}\rangle$. At this point, 
the description in terms of the four band original massive Dirac Hamiltonian can be substituted by a simpler description of the conduction 
band in the MCBB\cite{FuPRL2015}. 

\begin{figure}[t]
\begin{center}
\includegraphics[width=0.4\textwidth]{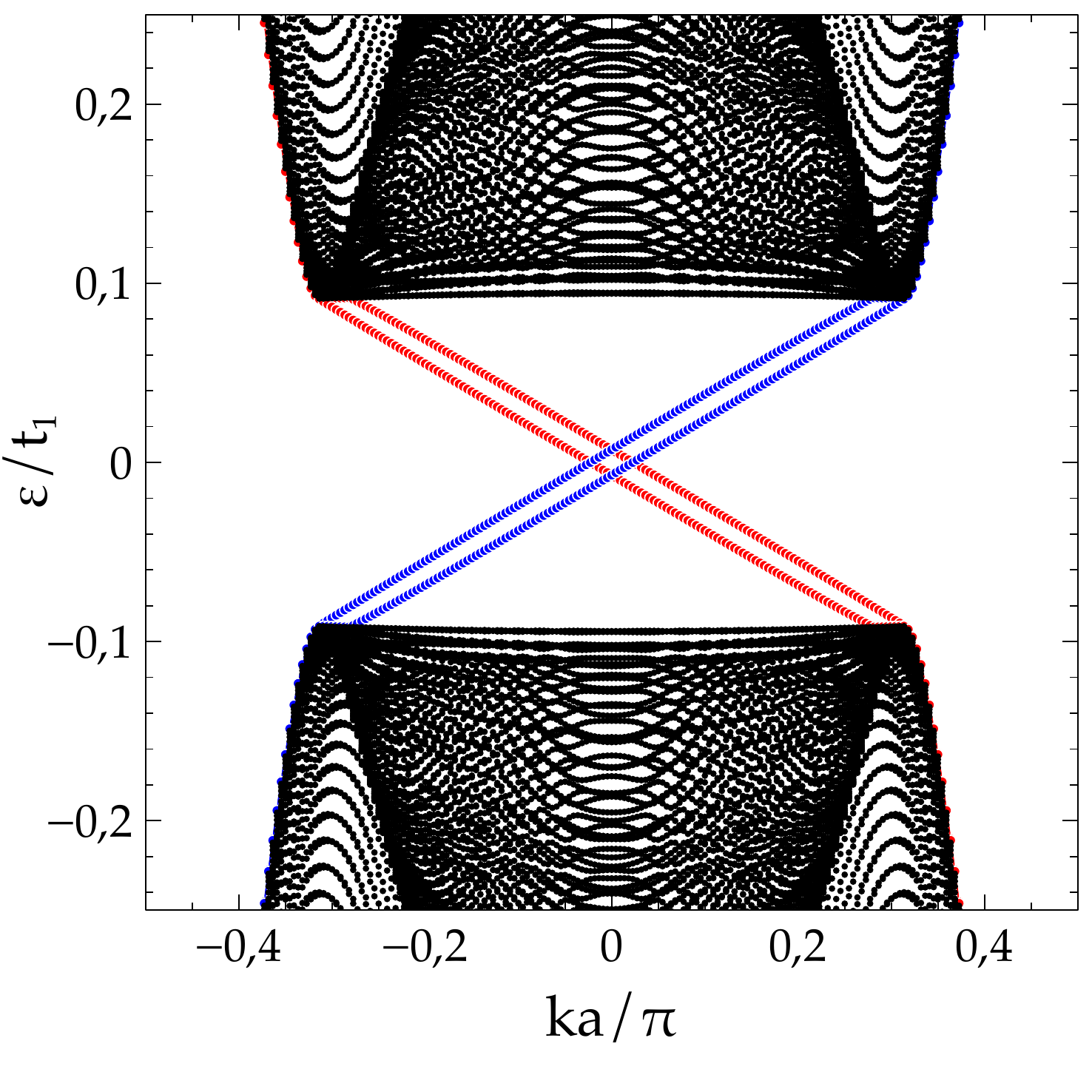}
\caption{Bands structure of a TI single layer with chiral superconductivity on a ribbon geometry. The parameters used are $t_0=-0.1$, 
$t_1=-5 t_2$, $t_2=1.0$, $\lambda=0.5$, $\Delta=0.1$, and $\mu=0.2$. Two Majorana edge states co-propagate on each side of the 
ribbon (blue on one side, red on the other side). 
\label{Fig:ChiralBands}}
\end{center}
\end{figure}

The BdG Hamiltonian Eq.~(\ref{Eq:BdG}) with the gap Eq.~(\ref{Eq:ChiralGap}) 
projected onto the conduction band reads
\begin{equation}\label{BdG4x4}
H_{\rm BdG}=\left(\begin{array}{cc}
\xi_{\bf k} & \Delta(k_x-ik_y)\tilde{s}_z\\
\Delta(k_x+ik_y)\tilde{s}_z & -\xi_{\bf k}
\end{array}\right).
\end{equation}
The system breaks TRI and belongs to class D topological superconductors in 2D. A finite Chern number $C_{\rm ch}=\pm 2$ is easily 
calculates from Eq.~(\ref{BdG4x4}), with the sign $\pm$ depending of whether the $(1,i)$ or $(1,-i)$ solution is realized. Correspondingly 
two copropagating chiral Majorana modes are expected to localize at the edge of the system. This can be seen directly from inspection of 
Eq.~(\ref{BdG4x4}). We see that both the spin up and spin down components are affected by a chiral $p_x+ip_y$ pairing, with gap of 
opposite sign for the two spin projections, due to the triplet nature of the pairing. Spinless chiral superconductivity in 2D opens a topologically 
non-trivial gap on the Fermi surface that gives rise to a single chiral Majorana mode at the boundary of the system, flowing with a velocity 
$v_M=\mp |\Delta|/k_F$, for the $p_x\pm ip_y$ cases, respectively. We then expect two chiral Majorana modes, each for spin component, 
that copropagate at the boundary of the system with $v_M=-|\Delta|/k_F$, with $\Delta$ the mean-field value of the order parameter $\psi_0$ 
in Eq.~(\ref{Eq:ChiralGap}). 

In order to check the predictions of the low energy effective model Eq.~(\ref{BdG4x4}) we calculate the bands of the tight-binding model 
on a ribbon geometry. The mean-field Hamiltonian is $H_{\rm MF}=H+H_{\rm sc}$, with $H_{\rm sc}$ the Hamiltonian term describing 
superconductivity at mean field level in the chiral phase
\begin{equation}
H_{\rm sc}=\sum_{i;\sigma s,\sigma' s'}c^\dag_{i,\sigma s}c^{\dag}_{i,\sigma' s'}\hat{\Delta}_{\sigma s;\sigma' s'}+{\rm H.c.},
\end{equation}
with $\hat{\Delta}=\Delta\sigma_y(s_x+is_y)is_y$. The band structure on a ribbon geometry is shown in Fig.~\ref{Fig:ChiralBands}, 
where clear co-propagating chiral Majorana modes appear, blue on one side and red on the other side of the ribbon.

\section{Conclusion}
\label{Sec:conclusions}

In this work we studied the stability of the chiral phase versus the nematic phase in Bi$_2$Se$_3$ as a function of the anisotropy of the 
system and the thickness of the sample. We showed that by increasing the two-dimensional character of the Fermi surface the chiral phase 
is expected to become stable. This can be experimentally achieved by properly choosing the doping element and their amount, that by 
intercalation modify the interlayer distance and enhance the two-dimensional character of the system.  A second root towards the realization 
of a chiral phase is achieved by exfoliation. We found that by reducing the number of layers constituting a thin slab of material the chiral phase 
can be stabilized already in few layer thick slabs and, for the low doping case, exfoliation down to $N<10$ layer is expected to show the chiral 
phase. In particular, the single layer case is shown to always favour the chiral phase. The resulting states is a chiral topological superconductor 
that hosts two copropagating chiral Majorana modes at its boundary. Our findings promote single layer of Bi$_2$Se$_3$ as an ideal material 
for the manifestation of chiral superconductivity, opening the route to topological quantum computations with Majorana modes.

\section{Acknowledgments}
The author acknowledges useful discussion with F. Finocchiaro, F. de Juan, and F. Guinea, and he is thankful to J. Schmalian for a 
key comment on the role of Rashba SOI. This work is supported by the European Union's Seventh Framework Programme (FP7/2007-2013) 
through the ERC Advanced Grant NOVGRAPHENE (GA No. 290846) and the Comunidad de Madrid through 
the grant MAD2D-CM, S2013/MIT-3007.

\bibliography{chiral-bib}

\begin{thebibliography}{54}
\expandafter\ifx\csname natexlab\endcsname\relax\def\natexlab#1{#1}\fi
\expandafter\ifx\csname bibnamefont\endcsname\relax
  \def\bibnamefont#1{#1}\fi
\expandafter\ifx\csname bibfnamefont\endcsname\relax
  \def\bibfnamefont#1{#1}\fi
\expandafter\ifx\csname citenamefont\endcsname\relax
  \def\citenamefont#1{#1}\fi
\expandafter\ifx\csname url\endcsname\relax
  \def\url#1{\texttt{#1}}\fi
\expandafter\ifx\csname urlprefix\endcsname\relax\def\urlprefix{URL }\fi
\providecommand{\bibinfo}[2]{#2}
\providecommand{\eprint}[2][]{\url{#2}}

\bibitem[{\citenamefont{Sigrist and Ueda}(1991)}]{SigristRMP1991}
\bibinfo{author}{\bibfnamefont{M.}~\bibnamefont{Sigrist}} \bibnamefont{and}
  \bibinfo{author}{\bibfnamefont{K.}~\bibnamefont{Ueda}},
  \bibinfo{journal}{Rev. Mod. Phys.} \textbf{\bibinfo{volume}{63}},
  \bibinfo{pages}{239} (\bibinfo{year}{1991}).

\bibitem[{\citenamefont{Qi and Zhang}(2011)}]{ZhangRMP2011}
\bibinfo{author}{\bibfnamefont{X.-L.} \bibnamefont{Qi}} \bibnamefont{and}
  \bibinfo{author}{\bibfnamefont{S.-C.} \bibnamefont{Zhang}},
  \bibinfo{journal}{Rev. Mod. Phys.} \textbf{\bibinfo{volume}{83}},
  \bibinfo{pages}{1057} (\bibinfo{year}{2011}).

\bibitem[{\citenamefont{Read and Green}(2000)}]{ReadGreen2000}
\bibinfo{author}{\bibfnamefont{N.}~\bibnamefont{Read}} \bibnamefont{and}
  \bibinfo{author}{\bibfnamefont{D.}~\bibnamefont{Green}},
  \bibinfo{journal}{Phys. Rev. B} \textbf{\bibinfo{volume}{61}},
  \bibinfo{pages}{10267} (\bibinfo{year}{2000}).

\bibitem[{\citenamefont{Ivanov}(2001)}]{IvanovPRL2001}
\bibinfo{author}{\bibfnamefont{D.~A.} \bibnamefont{Ivanov}},
  \bibinfo{journal}{Phys. Rev. Lett.} \textbf{\bibinfo{volume}{86}},
  \bibinfo{pages}{268} (\bibinfo{year}{2001}).

\bibitem[{\citenamefont{Alicea}(2012)}]{Alicea}
\bibinfo{author}{\bibfnamefont{J.}~\bibnamefont{Alicea}},
  \bibinfo{journal}{Reports on Progress in Physics}
  \textbf{\bibinfo{volume}{75}}, \bibinfo{pages}{076501}
  (\bibinfo{year}{2012}).

\bibitem[{\citenamefont{Beenakker}(2013)}]{Beenakker2013rev}
\bibinfo{author}{\bibfnamefont{C.~W.~J.} \bibnamefont{Beenakker}},
  \bibinfo{journal}{Annual Review of Condensed Matter Physics}
  \textbf{\bibinfo{volume}{4}}, \bibinfo{pages}{113} (\bibinfo{year}{2013}).

\bibitem[{\citenamefont{{Aguado}}(2017)}]{Aguado}
\bibinfo{author}{\bibfnamefont{R.}~\bibnamefont{{Aguado}}},
  \bibinfo{journal}{Riv. Nuovo Cimento} \textbf{\bibinfo{volume}{40}},
  \bibinfo{pages}{523} (\bibinfo{year}{2017}).

\bibitem[{\citenamefont{Meng and Balents}(2012)}]{MengPRB2012}
\bibinfo{author}{\bibfnamefont{T.}~\bibnamefont{Meng}} \bibnamefont{and}
  \bibinfo{author}{\bibfnamefont{L.}~\bibnamefont{Balents}},
  \bibinfo{journal}{Phys. Rev. B} \textbf{\bibinfo{volume}{86}},
  \bibinfo{pages}{054504} (\bibinfo{year}{2012}).

\bibitem[{\citenamefont{Sau and Tewari}(2012)}]{SauPRB2012}
\bibinfo{author}{\bibfnamefont{J.~D.} \bibnamefont{Sau}} \bibnamefont{and}
  \bibinfo{author}{\bibfnamefont{S.}~\bibnamefont{Tewari}},
  \bibinfo{journal}{Phys. Rev. B} \textbf{\bibinfo{volume}{86}},
  \bibinfo{pages}{104509} (\bibinfo{year}{2012}).

\bibitem[{\citenamefont{Yang et~al.}(2014)\citenamefont{Yang, Pan, and
  Zhang}}]{YangPRL2014}
\bibinfo{author}{\bibfnamefont{S.~A.} \bibnamefont{Yang}},
  \bibinfo{author}{\bibfnamefont{H.}~\bibnamefont{Pan}}, \bibnamefont{and}
  \bibinfo{author}{\bibfnamefont{F.}~\bibnamefont{Zhang}},
  \bibinfo{journal}{Phys. Rev. Lett.} \textbf{\bibinfo{volume}{113}},
  \bibinfo{pages}{046401} (\bibinfo{year}{2014}).

\bibitem[{\citenamefont{Biswas et~al.}(2013)\citenamefont{Biswas, Luetkens,
  Neupert, St\"urzer, Baines, Pascua, Schnyder, Fischer, Goryo, Lees
  et~al.}}]{Biswas2013}
\bibinfo{author}{\bibfnamefont{P.~K.} \bibnamefont{Biswas}},
  \bibinfo{author}{\bibfnamefont{H.}~\bibnamefont{Luetkens}},
  \bibinfo{author}{\bibfnamefont{T.}~\bibnamefont{Neupert}},
  \bibinfo{author}{\bibfnamefont{T.}~\bibnamefont{St\"urzer}},
  \bibinfo{author}{\bibfnamefont{C.}~\bibnamefont{Baines}},
  \bibinfo{author}{\bibfnamefont{G.}~\bibnamefont{Pascua}},
  \bibinfo{author}{\bibfnamefont{A.~P.} \bibnamefont{Schnyder}},
  \bibinfo{author}{\bibfnamefont{M.~H.} \bibnamefont{Fischer}},
  \bibinfo{author}{\bibfnamefont{J.}~\bibnamefont{Goryo}},
  \bibinfo{author}{\bibfnamefont{M.~R.} \bibnamefont{Lees}},
  \bibnamefont{et~al.}, \bibinfo{journal}{Phys. Rev. B}
  \textbf{\bibinfo{volume}{87}}, \bibinfo{pages}{180503}
  (\bibinfo{year}{2013}),
  \urlprefix\url{https://link.aps.org/doi/10.1103/PhysRevB.87.180503}.

\bibitem[{\citenamefont{Fischer et~al.}(2014)\citenamefont{Fischer, Neupert,
  Platt, Schnyder, Hanke, Goryo, Thomale, and Sigrist}}]{Fisher2014}
\bibinfo{author}{\bibfnamefont{M.~H.} \bibnamefont{Fischer}},
  \bibinfo{author}{\bibfnamefont{T.}~\bibnamefont{Neupert}},
  \bibinfo{author}{\bibfnamefont{C.}~\bibnamefont{Platt}},
  \bibinfo{author}{\bibfnamefont{A.~P.} \bibnamefont{Schnyder}},
  \bibinfo{author}{\bibfnamefont{W.}~\bibnamefont{Hanke}},
  \bibinfo{author}{\bibfnamefont{J.}~\bibnamefont{Goryo}},
  \bibinfo{author}{\bibfnamefont{R.}~\bibnamefont{Thomale}}, \bibnamefont{and}
  \bibinfo{author}{\bibfnamefont{M.}~\bibnamefont{Sigrist}},
  \bibinfo{journal}{Phys. Rev. B} \textbf{\bibinfo{volume}{89}},
  \bibinfo{pages}{020509} (\bibinfo{year}{2014}),
  \urlprefix\url{https://link.aps.org/doi/10.1103/PhysRevB.89.020509}.

\bibitem[{\citenamefont{Nayak et~al.}(2008)\citenamefont{Nayak, Simon, Stern,
  Freedman, and Das~Sarma}}]{NayakRMP2008}
\bibinfo{author}{\bibfnamefont{C.}~\bibnamefont{Nayak}},
  \bibinfo{author}{\bibfnamefont{S.~H.} \bibnamefont{Simon}},
  \bibinfo{author}{\bibfnamefont{A.}~\bibnamefont{Stern}},
  \bibinfo{author}{\bibfnamefont{M.}~\bibnamefont{Freedman}}, \bibnamefont{and}
  \bibinfo{author}{\bibfnamefont{S.}~\bibnamefont{Das~Sarma}},
  \bibinfo{journal}{Rev. Mod. Phys.} \textbf{\bibinfo{volume}{80}},
  \bibinfo{pages}{1083} (\bibinfo{year}{2008}).

\bibitem[{\citenamefont{Sarma et~al.}(2015)\citenamefont{Sarma, Freedman, and
  Nayak}}]{DasSarma2015}
\bibinfo{author}{\bibfnamefont{S.~D.} \bibnamefont{Sarma}},
  \bibinfo{author}{\bibfnamefont{M.}~\bibnamefont{Freedman}}, \bibnamefont{and}
  \bibinfo{author}{\bibfnamefont{C.}~\bibnamefont{Nayak}},
  \bibinfo{journal}{Npj Quantum Information} \textbf{\bibinfo{volume}{1}},
  \bibinfo{pages}{15001} (\bibinfo{year}{2015}).

\bibitem[{\citenamefont{Zhang et~al.}(2009)\citenamefont{Zhang, Liu, Qi, Dai,
  Fang, and Zhang}}]{ZhangNP2009}
\bibinfo{author}{\bibfnamefont{H.}~\bibnamefont{Zhang}},
  \bibinfo{author}{\bibfnamefont{C.-X.} \bibnamefont{Liu}},
  \bibinfo{author}{\bibfnamefont{X.-L.} \bibnamefont{Qi}},
  \bibinfo{author}{\bibfnamefont{X.}~\bibnamefont{Dai}},
  \bibinfo{author}{\bibfnamefont{Z.}~\bibnamefont{Fang}}, \bibnamefont{and}
  \bibinfo{author}{\bibfnamefont{S.-C.} \bibnamefont{Zhang}},
  \bibinfo{journal}{Nature Physics} \textbf{\bibinfo{volume}{5}},
  \bibinfo{pages}{438} (\bibinfo{year}{2009}).

\bibitem[{\citenamefont{Hasan and Kane}(2010)}]{HasanKaneRMP2010}
\bibinfo{author}{\bibfnamefont{M.~Z.} \bibnamefont{Hasan}} \bibnamefont{and}
  \bibinfo{author}{\bibfnamefont{C.~L.} \bibnamefont{Kane}},
  \bibinfo{journal}{Rev. Mod. Phys.} \textbf{\bibinfo{volume}{82}},
  \bibinfo{pages}{3045} (\bibinfo{year}{2010}).

\bibitem[{\citenamefont{Hor et~al.}(2010)\citenamefont{Hor, Williams,
  Checkelsky, Roushan, Seo, Xu, Zandbergen, Yazdani, Ong, and
  Cava}}]{HorPRL2010}
\bibinfo{author}{\bibfnamefont{Y.~S.} \bibnamefont{Hor}},
  \bibinfo{author}{\bibfnamefont{A.~J.} \bibnamefont{Williams}},
  \bibinfo{author}{\bibfnamefont{J.~G.} \bibnamefont{Checkelsky}},
  \bibinfo{author}{\bibfnamefont{P.}~\bibnamefont{Roushan}},
  \bibinfo{author}{\bibfnamefont{J.}~\bibnamefont{Seo}},
  \bibinfo{author}{\bibfnamefont{Q.}~\bibnamefont{Xu}},
  \bibinfo{author}{\bibfnamefont{H.~W.} \bibnamefont{Zandbergen}},
  \bibinfo{author}{\bibfnamefont{A.}~\bibnamefont{Yazdani}},
  \bibinfo{author}{\bibfnamefont{N.~P.} \bibnamefont{Ong}}, \bibnamefont{and}
  \bibinfo{author}{\bibfnamefont{R.~J.} \bibnamefont{Cava}},
  \bibinfo{journal}{Phys. Rev. Lett.} \textbf{\bibinfo{volume}{104}},
  \bibinfo{pages}{057001} (\bibinfo{year}{2010}).

\bibitem[{\citenamefont{Wray et~al.}(2010)\citenamefont{Wray, Xu, Xia, Hor,
  Qian, Fedorov, Lin, Bansil, Cava, and Hasan}}]{Wray2010}
\bibinfo{author}{\bibfnamefont{L.~A.} \bibnamefont{Wray}},
  \bibinfo{author}{\bibfnamefont{S.-Y.} \bibnamefont{Xu}},
  \bibinfo{author}{\bibfnamefont{Y.}~\bibnamefont{Xia}},
  \bibinfo{author}{\bibfnamefont{Y.~S.} \bibnamefont{Hor}},
  \bibinfo{author}{\bibfnamefont{D.}~\bibnamefont{Qian}},
  \bibinfo{author}{\bibfnamefont{A.~V.} \bibnamefont{Fedorov}},
  \bibinfo{author}{\bibfnamefont{H.}~\bibnamefont{Lin}},
  \bibinfo{author}{\bibfnamefont{A.}~\bibnamefont{Bansil}},
  \bibinfo{author}{\bibfnamefont{R.~J.} \bibnamefont{Cava}}, \bibnamefont{and}
  \bibinfo{author}{\bibfnamefont{M.~Z.} \bibnamefont{Hasan}},
  \bibinfo{journal}{Nature Physics} \textbf{\bibinfo{volume}{6}},
  \bibinfo{pages}{855} (\bibinfo{year}{2010}).

\bibitem[{\citenamefont{Kriener et~al.}(2011)\citenamefont{Kriener, Segawa,
  Ren, Sasaki, and Ando}}]{KrienerPRL2011}
\bibinfo{author}{\bibfnamefont{M.}~\bibnamefont{Kriener}},
  \bibinfo{author}{\bibfnamefont{K.}~\bibnamefont{Segawa}},
  \bibinfo{author}{\bibfnamefont{Z.}~\bibnamefont{Ren}},
  \bibinfo{author}{\bibfnamefont{S.}~\bibnamefont{Sasaki}}, \bibnamefont{and}
  \bibinfo{author}{\bibfnamefont{Y.}~\bibnamefont{Ando}},
  \bibinfo{journal}{Phys. Rev. Lett.} \textbf{\bibinfo{volume}{106}},
  \bibinfo{pages}{127004} (\bibinfo{year}{2011}).

\bibitem[{\citenamefont{Sasaki et~al.}(2011)\citenamefont{Sasaki, Kriener,
  Segawa, Yada, Tanaka, Sato, and Ando}}]{SasakiPRL2011}
\bibinfo{author}{\bibfnamefont{S.}~\bibnamefont{Sasaki}},
  \bibinfo{author}{\bibfnamefont{M.}~\bibnamefont{Kriener}},
  \bibinfo{author}{\bibfnamefont{K.}~\bibnamefont{Segawa}},
  \bibinfo{author}{\bibfnamefont{K.}~\bibnamefont{Yada}},
  \bibinfo{author}{\bibfnamefont{Y.}~\bibnamefont{Tanaka}},
  \bibinfo{author}{\bibfnamefont{M.}~\bibnamefont{Sato}}, \bibnamefont{and}
  \bibinfo{author}{\bibfnamefont{Y.}~\bibnamefont{Ando}},
  \bibinfo{journal}{Phys. Rev. Lett.} \textbf{\bibinfo{volume}{107}},
  \bibinfo{pages}{217001} (\bibinfo{year}{2011}).

\bibitem[{\citenamefont{Levy et~al.}(2013)\citenamefont{Levy, Zhang, Ha,
  Sharifi, Talin, Kuk, and Stroscio}}]{LevyPRL2013}
\bibinfo{author}{\bibfnamefont{N.}~\bibnamefont{Levy}},
  \bibinfo{author}{\bibfnamefont{T.}~\bibnamefont{Zhang}},
  \bibinfo{author}{\bibfnamefont{J.}~\bibnamefont{Ha}},
  \bibinfo{author}{\bibfnamefont{F.}~\bibnamefont{Sharifi}},
  \bibinfo{author}{\bibfnamefont{A.~A.} \bibnamefont{Talin}},
  \bibinfo{author}{\bibfnamefont{Y.}~\bibnamefont{Kuk}}, \bibnamefont{and}
  \bibinfo{author}{\bibfnamefont{J.~A.} \bibnamefont{Stroscio}},
  \bibinfo{journal}{Phys. Rev. Lett.} \textbf{\bibinfo{volume}{110}},
  \bibinfo{pages}{117001} (\bibinfo{year}{2013}).

\bibitem[{\citenamefont{Peng et~al.}(2013)\citenamefont{Peng, De, Lv, Wei, and
  Chu}}]{PengPRB2013}
\bibinfo{author}{\bibfnamefont{H.}~\bibnamefont{Peng}},
  \bibinfo{author}{\bibfnamefont{D.}~\bibnamefont{De}},
  \bibinfo{author}{\bibfnamefont{B.}~\bibnamefont{Lv}},
  \bibinfo{author}{\bibfnamefont{F.}~\bibnamefont{Wei}}, \bibnamefont{and}
  \bibinfo{author}{\bibfnamefont{C.-W.} \bibnamefont{Chu}},
  \bibinfo{journal}{Phys. Rev. B} \textbf{\bibinfo{volume}{88}},
  \bibinfo{pages}{024515} (\bibinfo{year}{2013}).

\bibitem[{\citenamefont{Fu and Berg}(2010)}]{FuBerg}
\bibinfo{author}{\bibfnamefont{L.}~\bibnamefont{Fu}} \bibnamefont{and}
  \bibinfo{author}{\bibfnamefont{E.}~\bibnamefont{Berg}},
  \bibinfo{journal}{Phys. Rev. Lett.} \textbf{\bibinfo{volume}{105}},
  \bibinfo{pages}{097001} (\bibinfo{year}{2010}).

\bibitem[{\citenamefont{Shruti et~al.}(2015)\citenamefont{Shruti, Maurya, Neha,
  Srivastava, and Patnaik}}]{shrutiPRB2015}
\bibinfo{author}{\bibnamefont{Shruti}}, \bibinfo{author}{\bibfnamefont{V.~K.}
  \bibnamefont{Maurya}},
  \bibinfo{author}{\bibfnamefont{P.}~\bibnamefont{Neha}},
  \bibinfo{author}{\bibfnamefont{P.}~\bibnamefont{Srivastava}},
  \bibnamefont{and} \bibinfo{author}{\bibfnamefont{S.}~\bibnamefont{Patnaik}},
  \bibinfo{journal}{Phys. Rev. B} \textbf{\bibinfo{volume}{92}},
  \bibinfo{pages}{020506} (\bibinfo{year}{2015}).

\bibitem[{\citenamefont{Liu et~al.}(2015)\citenamefont{Liu, Yao, Shao, Zuo, Pi,
  Tan, Zhang, and Zhang}}]{liuJACS2015}
\bibinfo{author}{\bibfnamefont{Z.}~\bibnamefont{Liu}},
  \bibinfo{author}{\bibfnamefont{X.}~\bibnamefont{Yao}},
  \bibinfo{author}{\bibfnamefont{J.}~\bibnamefont{Shao}},
  \bibinfo{author}{\bibfnamefont{M.}~\bibnamefont{Zuo}},
  \bibinfo{author}{\bibfnamefont{L.}~\bibnamefont{Pi}},
  \bibinfo{author}{\bibfnamefont{S.}~\bibnamefont{Tan}},
  \bibinfo{author}{\bibfnamefont{C.}~\bibnamefont{Zhang}}, \bibnamefont{and}
  \bibinfo{author}{\bibfnamefont{Y.}~\bibnamefont{Zhang}},
  \bibinfo{journal}{Journal of the American Chemical Society}
  \textbf{\bibinfo{volume}{137}}, \bibinfo{pages}{10512}
  (\bibinfo{year}{2015}).

\bibitem[{\citenamefont{Wang et~al.}(2016)\citenamefont{Wang, Taskin,
  Fr{\"o}lich, Braden, and Ando}}]{wangCM2016}
\bibinfo{author}{\bibfnamefont{Z.}~\bibnamefont{Wang}},
  \bibinfo{author}{\bibfnamefont{A.~A.} \bibnamefont{Taskin}},
  \bibinfo{author}{\bibfnamefont{T.}~\bibnamefont{Fr{\"o}lich}},
  \bibinfo{author}{\bibfnamefont{M.}~\bibnamefont{Braden}}, \bibnamefont{and}
  \bibinfo{author}{\bibfnamefont{Y.}~\bibnamefont{Ando}},
  \bibinfo{journal}{Chemistry of Materials} \textbf{\bibinfo{volume}{28}},
  \bibinfo{pages}{779} (\bibinfo{year}{2016}).

\bibitem[{\citenamefont{Asaba et~al.}(2017)\citenamefont{Asaba, Lawson,
  Tinsman, Chen, Corbae, Li, Qiu, Hor, Fu, and Li}}]{asabaPRX2017}
\bibinfo{author}{\bibfnamefont{T.}~\bibnamefont{Asaba}},
  \bibinfo{author}{\bibfnamefont{B.~J.} \bibnamefont{Lawson}},
  \bibinfo{author}{\bibfnamefont{C.}~\bibnamefont{Tinsman}},
  \bibinfo{author}{\bibfnamefont{L.}~\bibnamefont{Chen}},
  \bibinfo{author}{\bibfnamefont{P.}~\bibnamefont{Corbae}},
  \bibinfo{author}{\bibfnamefont{G.}~\bibnamefont{Li}},
  \bibinfo{author}{\bibfnamefont{Y.}~\bibnamefont{Qiu}},
  \bibinfo{author}{\bibfnamefont{Y.~S.} \bibnamefont{Hor}},
  \bibinfo{author}{\bibfnamefont{L.}~\bibnamefont{Fu}}, \bibnamefont{and}
  \bibinfo{author}{\bibfnamefont{L.}~\bibnamefont{Li}}, \bibinfo{journal}{Phys.
  Rev. X} \textbf{\bibinfo{volume}{7}}, \bibinfo{pages}{011009}
  (\bibinfo{year}{2017}).

\bibitem[{\citenamefont{Matano et~al.}(2016)\citenamefont{Matano, Kriener,
  Segawa, Ando, and Zheng}}]{Matano2016}
\bibinfo{author}{\bibfnamefont{K.}~\bibnamefont{Matano}},
  \bibinfo{author}{\bibfnamefont{M.}~\bibnamefont{Kriener}},
  \bibinfo{author}{\bibfnamefont{K.}~\bibnamefont{Segawa}},
  \bibinfo{author}{\bibfnamefont{Y.}~\bibnamefont{Ando}}, \bibnamefont{and}
  \bibinfo{author}{\bibfnamefont{G.-q.} \bibnamefont{Zheng}},
  \bibinfo{journal}{Nature Physics} \textbf{\bibinfo{volume}{12}},
  \bibinfo{pages}{852 EP} (\bibinfo{year}{2016}).

\bibitem[{\citenamefont{Yonezawa et~al.}(2016)\citenamefont{Yonezawa, Tajiri,
  Nakata, Nagai, Wang, Segawa, Ando, and Maeno}}]{YonezawaNP2016}
\bibinfo{author}{\bibfnamefont{S.}~\bibnamefont{Yonezawa}},
  \bibinfo{author}{\bibfnamefont{K.}~\bibnamefont{Tajiri}},
  \bibinfo{author}{\bibfnamefont{S.}~\bibnamefont{Nakata}},
  \bibinfo{author}{\bibfnamefont{Y.}~\bibnamefont{Nagai}},
  \bibinfo{author}{\bibfnamefont{Z.}~\bibnamefont{Wang}},
  \bibinfo{author}{\bibfnamefont{K.}~\bibnamefont{Segawa}},
  \bibinfo{author}{\bibfnamefont{Y.}~\bibnamefont{Ando}}, \bibnamefont{and}
  \bibinfo{author}{\bibfnamefont{Y.}~\bibnamefont{Maeno}},
  \bibinfo{journal}{Nature Physics} \textbf{\bibinfo{volume}{13}},
  \bibinfo{pages}{123 EP} (\bibinfo{year}{2016}).

\bibitem[{\citenamefont{Pan et~al.}(2016)\citenamefont{Pan, Nikitin, Araizi,
  Huang, Matsushita, Naka, and de~Visser}}]{PanSR2016}
\bibinfo{author}{\bibfnamefont{Y.}~\bibnamefont{Pan}},
  \bibinfo{author}{\bibfnamefont{A.~M.} \bibnamefont{Nikitin}},
  \bibinfo{author}{\bibfnamefont{G.~K.} \bibnamefont{Araizi}},
  \bibinfo{author}{\bibfnamefont{Y.~K.} \bibnamefont{Huang}},
  \bibinfo{author}{\bibfnamefont{Y.}~\bibnamefont{Matsushita}},
  \bibinfo{author}{\bibfnamefont{T.}~\bibnamefont{Naka}}, \bibnamefont{and}
  \bibinfo{author}{\bibfnamefont{A.}~\bibnamefont{de~Visser}},
  \bibinfo{journal}{Scientific Reports} \textbf{\bibinfo{volume}{6}},
  \bibinfo{pages}{28632 EP} (\bibinfo{year}{2016}).

\bibitem[{\citenamefont{Smylie et~al.}(2016)\citenamefont{Smylie, Claus, Welp,
  Kwok, Qiu, Hor, and Snezhko}}]{SmyliePRB2016}
\bibinfo{author}{\bibfnamefont{M.~P.} \bibnamefont{Smylie}},
  \bibinfo{author}{\bibfnamefont{H.}~\bibnamefont{Claus}},
  \bibinfo{author}{\bibfnamefont{U.}~\bibnamefont{Welp}},
  \bibinfo{author}{\bibfnamefont{W.-K.} \bibnamefont{Kwok}},
  \bibinfo{author}{\bibfnamefont{Y.}~\bibnamefont{Qiu}},
  \bibinfo{author}{\bibfnamefont{Y.~S.} \bibnamefont{Hor}}, \bibnamefont{and}
  \bibinfo{author}{\bibfnamefont{A.}~\bibnamefont{Snezhko}},
  \bibinfo{journal}{Phys. Rev. B} \textbf{\bibinfo{volume}{94}},
  \bibinfo{pages}{180510} (\bibinfo{year}{2016}).

\bibitem[{\citenamefont{Shen et~al.}(2017)\citenamefont{Shen, He, Yuan, Huang,
  Cho, Lee, Hor, Law, and Lortz}}]{Shen2017}
\bibinfo{author}{\bibfnamefont{J.}~\bibnamefont{Shen}},
  \bibinfo{author}{\bibfnamefont{W.-Y.} \bibnamefont{He}},
  \bibinfo{author}{\bibfnamefont{N.~F.~Q.} \bibnamefont{Yuan}},
  \bibinfo{author}{\bibfnamefont{Z.}~\bibnamefont{Huang}},
  \bibinfo{author}{\bibfnamefont{C.-w.} \bibnamefont{Cho}},
  \bibinfo{author}{\bibfnamefont{S.~H.} \bibnamefont{Lee}},
  \bibinfo{author}{\bibfnamefont{Y.~S.} \bibnamefont{Hor}},
  \bibinfo{author}{\bibfnamefont{K.~T.} \bibnamefont{Law}}, \bibnamefont{and}
  \bibinfo{author}{\bibfnamefont{R.}~\bibnamefont{Lortz}},
  \bibinfo{journal}{npj Quantum Materials} \textbf{\bibinfo{volume}{2}},
  \bibinfo{pages}{59} (\bibinfo{year}{2017}).

\bibitem[{\citenamefont{Fu}(2014)}]{FuPRBR2014}
\bibinfo{author}{\bibfnamefont{L.}~\bibnamefont{Fu}}, \bibinfo{journal}{Phys.
  Rev. B} \textbf{\bibinfo{volume}{90}}, \bibinfo{pages}{100509}
  (\bibinfo{year}{2014}).

\bibitem[{\citenamefont{Venderbos
  et~al.}(2016{\natexlab{a}})\citenamefont{Venderbos, Kozii, and
  Fu}}]{Venderbos2015}
\bibinfo{author}{\bibfnamefont{J.~W.~F.} \bibnamefont{Venderbos}},
  \bibinfo{author}{\bibfnamefont{V.}~\bibnamefont{Kozii}}, \bibnamefont{and}
  \bibinfo{author}{\bibfnamefont{L.}~\bibnamefont{Fu}}, \bibinfo{journal}{Phys.
  Rev. B} \textbf{\bibinfo{volume}{94}}, \bibinfo{pages}{180504}
  (\bibinfo{year}{2016}{\natexlab{a}}).

\bibitem[{\citenamefont{Hashimoto et~al.}(2013)\citenamefont{Hashimoto, Yada,
  Yamakage, Sato, and Tanaka}}]{HashimotoJPSJ2013}
\bibinfo{author}{\bibfnamefont{T.}~\bibnamefont{Hashimoto}},
  \bibinfo{author}{\bibfnamefont{K.}~\bibnamefont{Yada}},
  \bibinfo{author}{\bibfnamefont{A.}~\bibnamefont{Yamakage}},
  \bibinfo{author}{\bibfnamefont{M.}~\bibnamefont{Sato}}, \bibnamefont{and}
  \bibinfo{author}{\bibfnamefont{Y.}~\bibnamefont{Tanaka}},
  \bibinfo{journal}{Journal of the Physical Society of Japan}
  \textbf{\bibinfo{volume}{82}}, \bibinfo{pages}{044704}
  (\bibinfo{year}{2013}).

\bibitem[{\citenamefont{Nagai and Ota}(2016)}]{NagaiPRB2016}
\bibinfo{author}{\bibfnamefont{Y.}~\bibnamefont{Nagai}} \bibnamefont{and}
  \bibinfo{author}{\bibfnamefont{Y.}~\bibnamefont{Ota}},
  \bibinfo{journal}{Phys. Rev. B} \textbf{\bibinfo{volume}{94}},
  \bibinfo{pages}{134516} (\bibinfo{year}{2016}).

\bibitem[{\citenamefont{Venderbos
  et~al.}(2016{\natexlab{b}})\citenamefont{Venderbos, Kozii, and
  Fu}}]{VenderbosPRB2016-2}
\bibinfo{author}{\bibfnamefont{J.~W.~F.} \bibnamefont{Venderbos}},
  \bibinfo{author}{\bibfnamefont{V.}~\bibnamefont{Kozii}}, \bibnamefont{and}
  \bibinfo{author}{\bibfnamefont{L.}~\bibnamefont{Fu}}, \bibinfo{journal}{Phys.
  Rev. B} \textbf{\bibinfo{volume}{94}}, \bibinfo{pages}{094522}
  (\bibinfo{year}{2016}{\natexlab{b}}).

\bibitem[{\citenamefont{Wu and
  Martin}(2017{\natexlab{a}})}]{FengchengPRB2017-1}
\bibinfo{author}{\bibfnamefont{F.}~\bibnamefont{Wu}} \bibnamefont{and}
  \bibinfo{author}{\bibfnamefont{I.}~\bibnamefont{Martin}},
  \bibinfo{journal}{Phys. Rev. B} \textbf{\bibinfo{volume}{95}},
  \bibinfo{pages}{224503} (\bibinfo{year}{2017}{\natexlab{a}}).

\bibitem[{\citenamefont{Zyuzin et~al.}(2017)\citenamefont{Zyuzin, Garaud, and
  Babaev}}]{ZyuzinPRL2017}
\bibinfo{author}{\bibfnamefont{A.~A.} \bibnamefont{Zyuzin}},
  \bibinfo{author}{\bibfnamefont{J.}~\bibnamefont{Garaud}}, \bibnamefont{and}
  \bibinfo{author}{\bibfnamefont{E.}~\bibnamefont{Babaev}},
  \bibinfo{journal}{Phys. Rev. Lett.} \textbf{\bibinfo{volume}{119}},
  \bibinfo{pages}{167001} (\bibinfo{year}{2017}).

\bibitem[{\citenamefont{Chirolli et~al.}(2017)\citenamefont{Chirolli, de~Juan,
  and Guinea}}]{ChirolliPRB2017}
\bibinfo{author}{\bibfnamefont{L.}~\bibnamefont{Chirolli}},
  \bibinfo{author}{\bibfnamefont{F.}~\bibnamefont{de~Juan}}, \bibnamefont{and}
  \bibinfo{author}{\bibfnamefont{F.}~\bibnamefont{Guinea}},
  \bibinfo{journal}{Phys. Rev. B} \textbf{\bibinfo{volume}{95}},
  \bibinfo{pages}{201110} (\bibinfo{year}{2017}).

\bibitem[{\citenamefont{Yuan et~al.}(2017)\citenamefont{Yuan, He, and
  Law}}]{NoahPRB2017}
\bibinfo{author}{\bibfnamefont{N.~F.~Q.} \bibnamefont{Yuan}},
  \bibinfo{author}{\bibfnamefont{W.-Y.} \bibnamefont{He}}, \bibnamefont{and}
  \bibinfo{author}{\bibfnamefont{K.~T.} \bibnamefont{Law}},
  \bibinfo{journal}{Phys. Rev. B} \textbf{\bibinfo{volume}{95}},
  \bibinfo{pages}{201109} (\bibinfo{year}{2017}).

\bibitem[{\citenamefont{Wu and Martin}(2017{\natexlab{b}})}]{FengchengPRB2017}
\bibinfo{author}{\bibfnamefont{F.}~\bibnamefont{Wu}} \bibnamefont{and}
  \bibinfo{author}{\bibfnamefont{I.}~\bibnamefont{Martin}},
  \bibinfo{journal}{Phys. Rev. B} \textbf{\bibinfo{volume}{96}},
  \bibinfo{pages}{144504} (\bibinfo{year}{2017}{\natexlab{b}}).

\bibitem[{\citenamefont{Fernandes et~al.}(2012)\citenamefont{Fernandes,
  Chubukov, Knolle, Eremin, and Schmalian}}]{Fernandes2012}
\bibinfo{author}{\bibfnamefont{R.~M.} \bibnamefont{Fernandes}},
  \bibinfo{author}{\bibfnamefont{A.~V.} \bibnamefont{Chubukov}},
  \bibinfo{author}{\bibfnamefont{J.}~\bibnamefont{Knolle}},
  \bibinfo{author}{\bibfnamefont{I.}~\bibnamefont{Eremin}}, \bibnamefont{and}
  \bibinfo{author}{\bibfnamefont{J.}~\bibnamefont{Schmalian}},
  \bibinfo{journal}{Phys. Rev. B} \textbf{\bibinfo{volume}{85}},
  \bibinfo{pages}{024534} (\bibinfo{year}{2012}).

\bibitem[{\citenamefont{{Hecker} and {Schmalian}}(2018)}]{Schmalian2017}
\bibinfo{author}{\bibfnamefont{M.}~\bibnamefont{{Hecker}}} \bibnamefont{and}
  \bibinfo{author}{\bibfnamefont{J.}~\bibnamefont{{Schmalian}}},
  \bibinfo{journal}{npj Quantum Materials} \textbf{\bibinfo{volume}{3}},
  \bibinfo{pages}{26} (\bibinfo{year}{2018}).

\bibitem[{\citenamefont{Zhang et~al.}(2010)\citenamefont{Zhang, He, Chang,
  Song, Wang, Chen, Jia, Fang, Dai, Shan et~al.}}]{ZhangNP2010crossover}
\bibinfo{author}{\bibfnamefont{Y.}~\bibnamefont{Zhang}},
  \bibinfo{author}{\bibfnamefont{K.}~\bibnamefont{He}},
  \bibinfo{author}{\bibfnamefont{C.-Z.} \bibnamefont{Chang}},
  \bibinfo{author}{\bibfnamefont{C.-L.} \bibnamefont{Song}},
  \bibinfo{author}{\bibfnamefont{L.-L.} \bibnamefont{Wang}},
  \bibinfo{author}{\bibfnamefont{X.}~\bibnamefont{Chen}},
  \bibinfo{author}{\bibfnamefont{J.-F.} \bibnamefont{Jia}},
  \bibinfo{author}{\bibfnamefont{Z.}~\bibnamefont{Fang}},
  \bibinfo{author}{\bibfnamefont{X.}~\bibnamefont{Dai}},
  \bibinfo{author}{\bibfnamefont{W.-Y.} \bibnamefont{Shan}},
  \bibnamefont{et~al.}, \bibinfo{journal}{Nature Physics}
  \textbf{\bibinfo{volume}{6}}, \bibinfo{pages}{712 EP }
  (\bibinfo{year}{2010}).

\bibitem[{\citenamefont{Zhang et~al.}(2011)\citenamefont{Zhang, Qin, Chen, He,
  Lu, Li, and Wu}}]{ZhangAFM2011}
\bibinfo{author}{\bibfnamefont{G.}~\bibnamefont{Zhang}},
  \bibinfo{author}{\bibfnamefont{H.}~\bibnamefont{Qin}},
  \bibinfo{author}{\bibfnamefont{J.}~\bibnamefont{Chen}},
  \bibinfo{author}{\bibfnamefont{X.}~\bibnamefont{He}},
  \bibinfo{author}{\bibfnamefont{L.}~\bibnamefont{Lu}},
  \bibinfo{author}{\bibfnamefont{Y.}~\bibnamefont{Li}}, \bibnamefont{and}
  \bibinfo{author}{\bibfnamefont{K.}~\bibnamefont{Wu}},
  \bibinfo{journal}{Advanced Functional Materials}
  \textbf{\bibinfo{volume}{21}}, \bibinfo{pages}{2351} (\bibinfo{year}{2011}).

\bibitem[{\citenamefont{Scheurer et~al.}(2017)\citenamefont{Scheurer,
  Agterberg, and Schmalian}}]{Scheurer2017}
\bibinfo{author}{\bibfnamefont{M.~S.} \bibnamefont{Scheurer}},
  \bibinfo{author}{\bibfnamefont{D.~F.} \bibnamefont{Agterberg}},
  \bibnamefont{and}
  \bibinfo{author}{\bibfnamefont{J.}~\bibnamefont{Schmalian}},
  \bibinfo{journal}{npj Quantum Materials} \textbf{\bibinfo{volume}{2}},
  \bibinfo{pages}{9} (\bibinfo{year}{2017}).

\bibitem[{\citenamefont{Schnyder et~al.}(2008)\citenamefont{Schnyder, Ryu,
  Furusaki, and Ludwig}}]{Schnyder}
\bibinfo{author}{\bibfnamefont{A.~P.} \bibnamefont{Schnyder}},
  \bibinfo{author}{\bibfnamefont{S.}~\bibnamefont{Ryu}},
  \bibinfo{author}{\bibfnamefont{A.}~\bibnamefont{Furusaki}}, \bibnamefont{and}
  \bibinfo{author}{\bibfnamefont{A.~W.~W.} \bibnamefont{Ludwig}},
  \bibinfo{journal}{Phys. Rev. B} \textbf{\bibinfo{volume}{78}},
  \bibinfo{pages}{195125} (\bibinfo{year}{2008}),
  \urlprefix\url{https://link.aps.org/doi/10.1103/PhysRevB.78.195125}.

\bibitem[{\citenamefont{Chirolli et~al.}(2018)\citenamefont{Chirolli,
  Baltan\'as, and Frustaglia}}]{Chirolli2018}
\bibinfo{author}{\bibfnamefont{L.}~\bibnamefont{Chirolli}},
  \bibinfo{author}{\bibfnamefont{J.~P.} \bibnamefont{Baltan\'as}},
  \bibnamefont{and}
  \bibinfo{author}{\bibfnamefont{D.}~\bibnamefont{Frustaglia}},
  \bibinfo{journal}{Phys. Rev. B} \textbf{\bibinfo{volume}{97}},
  \bibinfo{pages}{155416} (\bibinfo{year}{2018}),
  \urlprefix\url{https://link.aps.org/doi/10.1103/PhysRevB.97.155416}.

\bibitem[{\citenamefont{Parhizgar and Black-Schaffer}(2017)}]{Parhizgar2017}
\bibinfo{author}{\bibfnamefont{F.}~\bibnamefont{Parhizgar}} \bibnamefont{and}
  \bibinfo{author}{\bibfnamefont{A.~M.} \bibnamefont{Black-Schaffer}},
  \bibinfo{journal}{Scientific Reports} \textbf{\bibinfo{volume}{7}},
  \bibinfo{pages}{9817} (\bibinfo{year}{2017}).

\bibitem[{\citenamefont{Nakosai et~al.}(2012)\citenamefont{Nakosai, Tanaka, and
  Nagaosa}}]{Nakosai-PRL2012}
\bibinfo{author}{\bibfnamefont{S.}~\bibnamefont{Nakosai}},
  \bibinfo{author}{\bibfnamefont{Y.}~\bibnamefont{Tanaka}}, \bibnamefont{and}
  \bibinfo{author}{\bibfnamefont{N.}~\bibnamefont{Nagaosa}},
  \bibinfo{journal}{Phys. Rev. Lett.} \textbf{\bibinfo{volume}{108}},
  \bibinfo{pages}{147003} (\bibinfo{year}{2012}).

\bibitem[{\citenamefont{Fu}(2015)}]{FuPRL2015}
\bibinfo{author}{\bibfnamefont{L.}~\bibnamefont{Fu}}, \bibinfo{journal}{Phys.
  Rev. Lett.} \textbf{\bibinfo{volume}{115}}, \bibinfo{pages}{026401}
  (\bibinfo{year}{2015}).

\bibitem[{\citenamefont{Hsieh and Fu}(2012)}]{HsiehFu-PRL2012}
\bibinfo{author}{\bibfnamefont{T.~H.} \bibnamefont{Hsieh}} \bibnamefont{and}
  \bibinfo{author}{\bibfnamefont{L.}~\bibnamefont{Fu}}, \bibinfo{journal}{Phys.
  Rev. Lett.} \textbf{\bibinfo{volume}{108}}, \bibinfo{pages}{107005}
  (\bibinfo{year}{2012}).

\bibitem[{\citenamefont{Lahoud et~al.}(2013)\citenamefont{Lahoud, Maniv,
  Petrushevsky, Naamneh, Ribak, Wiedmann, Petaccia, Salman, Chashka, Dagan
  et~al.}}]{LahoudPRB2013}
\bibinfo{author}{\bibfnamefont{E.}~\bibnamefont{Lahoud}},
  \bibinfo{author}{\bibfnamefont{E.}~\bibnamefont{Maniv}},
  \bibinfo{author}{\bibfnamefont{M.~S.} \bibnamefont{Petrushevsky}},
  \bibinfo{author}{\bibfnamefont{M.}~\bibnamefont{Naamneh}},
  \bibinfo{author}{\bibfnamefont{A.}~\bibnamefont{Ribak}},
  \bibinfo{author}{\bibfnamefont{S.}~\bibnamefont{Wiedmann}},
  \bibinfo{author}{\bibfnamefont{L.}~\bibnamefont{Petaccia}},
  \bibinfo{author}{\bibfnamefont{Z.}~\bibnamefont{Salman}},
  \bibinfo{author}{\bibfnamefont{K.~B.} \bibnamefont{Chashka}},
  \bibinfo{author}{\bibfnamefont{Y.}~\bibnamefont{Dagan}},
  \bibnamefont{et~al.}, \bibinfo{journal}{Phys. Rev. B}
  \textbf{\bibinfo{volume}{88}}, \bibinfo{pages}{195107}
  (\bibinfo{year}{2013}).

\end{thebibliography}

\end{document}